\documentclass{sf2a-conf2019}
\usepackage{graphicx}
\usepackage{hyperref}
\usepackage[]{natbib}  
\usepackage{epstopdf}
\usepackage{wrapfig}
\usepackage{bm}

\def\BibTeX{{\rm B\kern-.05em{\sc i\kern-.025em b}\kern-.08em
    T\kern-.1667em\lower.7ex\hbox{E}\kern-.125emX}}
\bibpunct{(}{)}{;}{a}{}{,}  

\begin{document}

\TitreGlobal{SF2A 2019}


\title{Determining surface rotation periods of solar-like stars observed
by the Kepler mission using Machine Learning techniques}

\runningtitle{Machine learning analysis of MS stars rotation}

\author{S.N.~Breton$^{1,}$}\address{IRFU, CEA, Université Paris-Saclay, 91191 Gif-sur-Yvette, France} \address{AIM, CEA, CNRS, Université Paris-Saclay, Université Paris Diderot, Sorbonne Paris Cité, 91191 Gif-sur-Yvette, France}

\author{L.~Bugnet$^{1,2}$}

\author{A.R.G.~Santos}\address{Space Science Institute, 4750 Walnut Street Suite 205, Boulder, CO 80301, USA}

\author{A.~Le~Saux$^{1,2}$}

\author{S.~Mathur$^{4,}$}\address{Instituto de Astrofísica de Canarias, 38200 La Laguna, Tenerife, Spain}\address{Universidad de La Laguna, Dpto. de Astrofísica, 38205 La Laguna, Tenerife, Spain}

\author{P.L.~Pall\'{e}$^{4,5}$}

\author{R.A.~Garc\'{i}a$^{1,2}$}




\setcounter{page}{237}


\maketitle


\begin{abstract}
For a solar-like star, the surface rotation evolves with time, allowing in principle to estimate the age of a star from its surface rotation period. Here we are interested in measuring surface rotation periods of solar-like stars observed by the NASA mission \textit{Kepler}. Different methods have been developed to track rotation signals in \textit{Kepler} photometric light curves: time-frequency analysis based on wavelet techniques, autocorrelation and composite spectrum. 
We use the learning abilities of random forest classifiers to take decisions during two crucial steps of the analysis. First, given some input parameters, we discriminate the considered \textit{Kepler} targets between rotating MS stars, non-rotating MS stars, red giants, binaries and pulsators. We then use a second classifier only on the MS rotating targets to decide the best data-analysis treatment. 
\end{abstract}



\section{Introduction}

Rotation plays an important role in stellar evolution. For cool main-sequence (MS) dwarfs (G, K and M spectral type), age may be determined thanks to gyrochronology \citep{1972ApJ...171..565S}: surface rotation evolves roughly as the square root of its age. Even if recent studies suggested that at a given stage of its evolution, the braking of a solar-like star is reduced \citep{2016Natur.529..181V}, this relation seems to be verified while the stars remain on the main sequence.

In this work, we use \textit{Kepler} \citep{2010Sci...327..977B} photometric light curves obtained with the KADACS pipeline (\textit{Kepler} Asteroseismic Data Analysis and Calibration Software, \citealt{2011MNRAS.414L...6G,2014A&A...568A..10G}). The KADACS pipeline has been specifically designed to correct \textit{Kepler} light curves from instrumental effects and properly stitch the quarters in an optimized way for asteroseismology studies. We consider different high-pass filters (20, 55, and 80 days) for the processing of the light curves to be sure that rotation period is not filtered out. Rotation period is then extracted thanks to a combination of different methods (Global Wavelet Power Spectrum GWPS, AutoCorrelation Function ACF and Composite Spectrum CS) as described in \citet{2010A&A...511A..46M}, \citet{2014A&A...572A..34G}, \citet{2016MNRAS.456..119C} and \citet{2017A&A...605A.111C}.
However, the computed rotation period may differ from one KADACS filter to another and from one method to another. Dozens of thousands of stars have to be considered with, until now, no other solution than to use a pre-defined hierarchical decision tree and make a final visual inspection of the conflicting cases. A machine learning algorithm seems the ideal tool to make the decision over the stars of our data set. Before performing this analysis and to avoid any bias, it is nevertheless necessary to distinguish main sequence rotators from other types of targets. 

\begin{wrapfigure}{L}{0.4 \textwidth}
    \centering
    \includegraphics[width = 0.39 \textwidth]{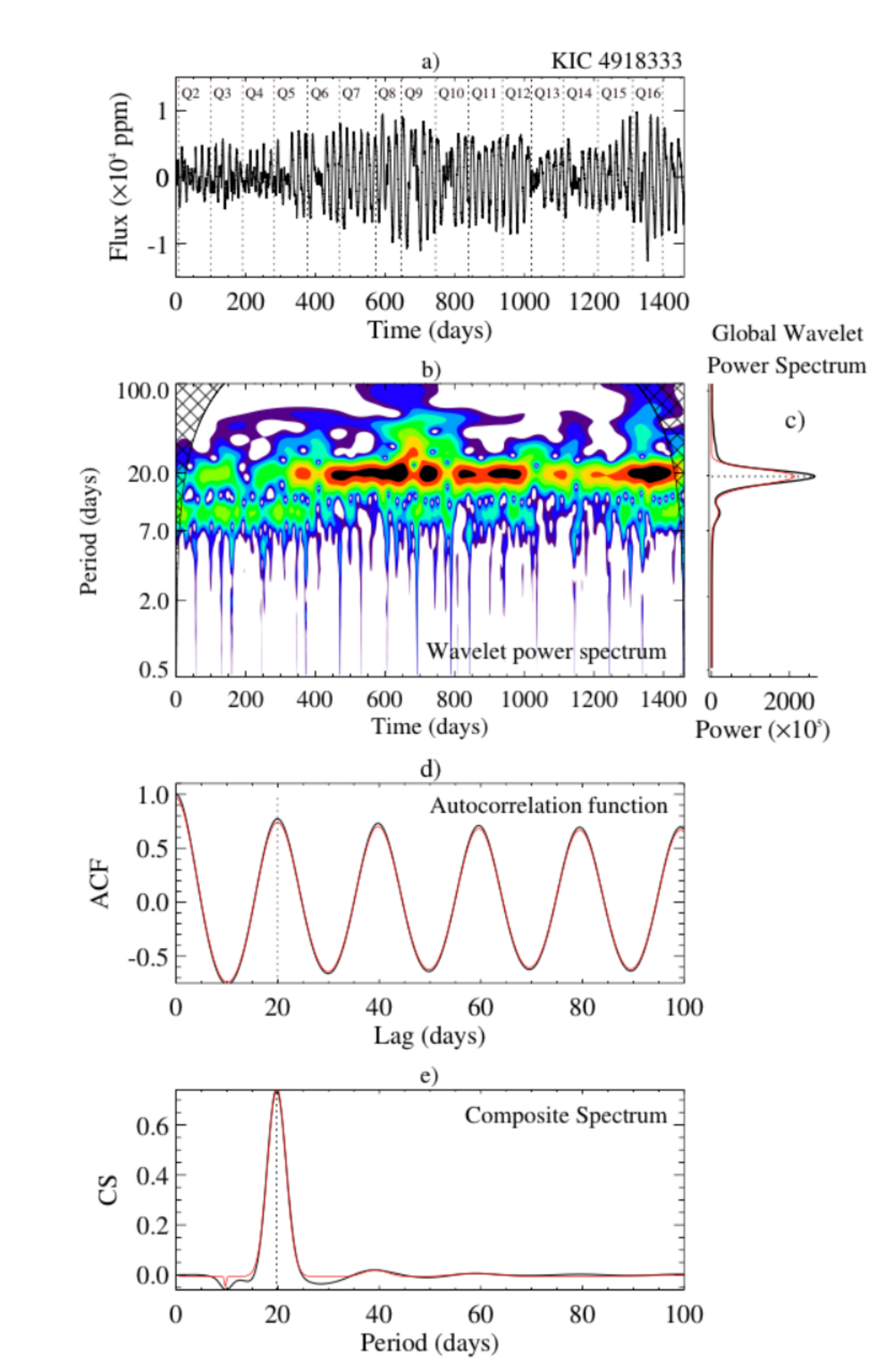}
    \caption{\textit{From top to bottom} - example of \textit{Kepler} photometric lightcurve analyzed with A2Z pipeline \citep{2010A&A...511A..46M}, wavelet power spectrum, autocorrelation function and composite spectrum. Extracted from \citet{santosAPJinprep}.}
    \label{fig:period_extraction}
    \vspace{-65pt}
\end{wrapfigure}
  
\section{Rotators classification}

Our set of stars consist of 14,441 M and K dwarfs based on the \textit{Kepler} star properties catalog from \citet{2017ApJS..229...30M}, whose rotation periods have been studied by \citet{santosAPJinprep}. However, the sample can be polluted by red giants (RG), classical pulsators (CP) or eclipsing binaries. \citeauthor{santosAPJinprep} identified these pollutors by visually checking the light curves. Here, we propose to use artificial intelligence methods to automatically detect such pollutions. We train a first random forest algorithm (with the Python package \textit{scikit-learn}, \citealt{scikit-learn})  in order to identify those different targets. The principle of a random forest algorithm is briefly reminded in annex A1.
The input parameters are:
\begin{itemize}
    \item periods computed by each method, $P_\mathrm{GWPS}$, $P_\mathrm{ACF}$, $P_\mathrm{CS}$, and related control values $H_\mathrm{ACF}$, $G_\mathrm{ACF}$ and $H_\mathrm{CS}$ (see Figure~\ref{fig:period_extraction} and \citealt{2017A&A...605A.111C} for further explanation);
    \item photometric activity proxy S$_\mathrm{ph}$ \citep{2010arXiv1012.0494G,2014A&A...572A..34G,2014A&A...562A.124M}; 
    \item FliPer values (see \citealt{2018A&A...620A..38B,2019A&A...624A..79B}); 
    \item effective temperature $T_{\mathrm{eff}}$ and surface gravity $\log g$ from \citet{2017ApJS..229...30M}.
\end{itemize}
    
\begin{figure}[b!]
    \centering
    \includegraphics[width = 1. \textwidth]{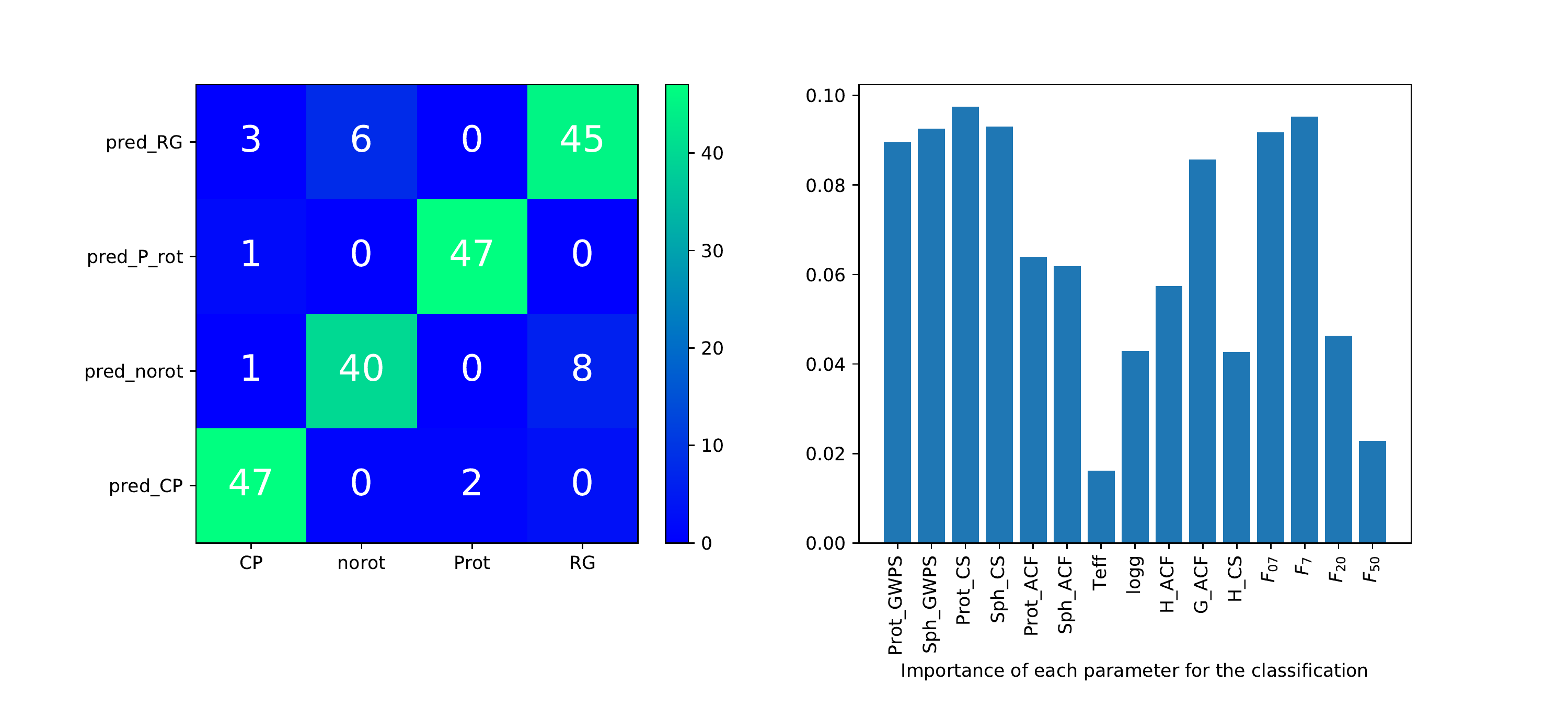}
    \caption{\textit{Left panel} - classification result for a test set of 200 stars. The algorithm has been trained with 600 stars that were visually classified before-hand. CP stands for classical pulsator, norot for MS non-rotating star, Prot for MS rotating star, RG for red giant. The real class of an element corresponds to its column label, the class assignated by the classifier corresponds to the line label. Accuracy of the classification is 0.895. \textit{Right panel} : relative importance of each parameter used for the classification.}
    \label{fig:first_rf}
\end{figure}    
    
Those parameters have been chosen because their values are directly related to stellar types and rotation properties. $T_{\mathrm{eff}}$ and $\log g$ are good parameters to distinguish between red giants and MS stars. The FliPer metric allows us to help disentangling the proposed classes of stars attending to their power in the PSD. Rotation periods computed by the three methods combined with all the other related parameters ($H_\mathrm{ACF}$, $G_\mathrm{ACF}$ and $H_\mathrm{CS}$) allow the classifier to decide whether the measured signal corresponds to a rotation period or not. Figure~\ref{fig:first_rf} shows the result of the classification of the test set. Stars are globally well classified, except for some non-rotating main-sequence stars and red giants with close $T_{\mathrm{eff}}$ and $\log g$. On a $T_{\mathrm{eff}}$-$\log g$ diagram, those stars would lay in the subgiants region. Thus, one of the next improvements of the classifier will be to add the possibility for the algorithm to give a label \textit{subgiant}.

\section{Period determination}
 
\begin{figure}[h!]
    \centering
    \includegraphics[width = 1. \textwidth]{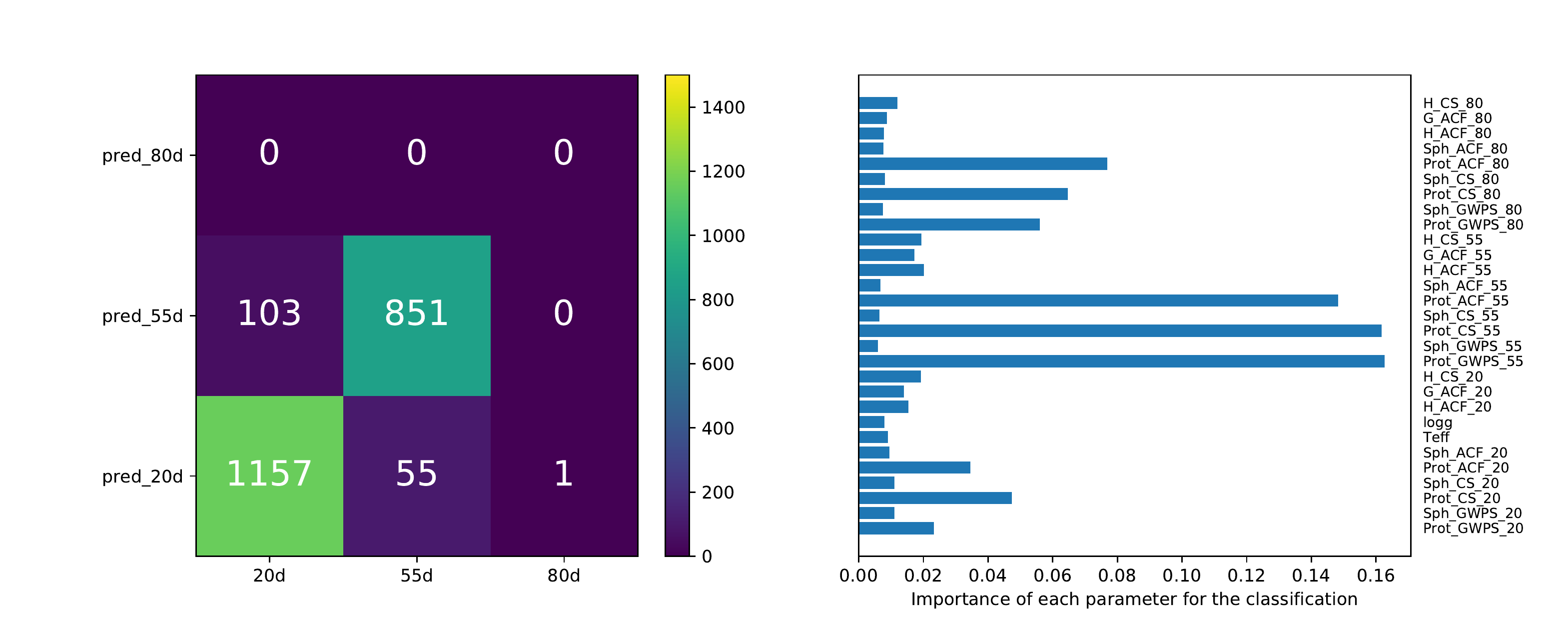}
    \caption{\textit{Left panel}: filter-choice result for a test set of 2166 rotating MS stars. The algorithm was trained with 12,275 stars. The total number of stars is 14,441. Classifier accuracy is 0.927. The «true» accuracy on the period is 0.974. \textit{Right panel}: relative importance of each parameter used for the classification. $P_\mathrm{GWPS}$, $P_\mathrm{ACF}$, $P_\mathrm{CS}$, $H_\mathrm{ACF}$, $G_\mathrm{ACF}$, $H_\mathrm{CS}$ and S$_\mathrm{ph}$ values are considered for each filter (20, 55, 80) and consequently subscripted in the legend of the plot.}
    \label{fig:second_rf}
\end{figure}

A second random forest classifier is trained to determine the best filter to consider (20, 55, 80 days) to retrieve the most probable rotation period $P_\mathrm{rot}$. We assume that this $P_\mathrm{rot}$ will be given by the wavelet method of the correct filter. Comparing the best filter choice and the classifier choice gives us an estimation of the classifier accuracy.
Sometimes, even when the classifier choice is not the filter chosen by \citet{santosAPJinprep}, the period estimate is approximately the same. If the period differs by less than 10\% from the true period labelled on the training set, we consider that the classifier is right and compute what we call the true accuracy (e.g. the true accuracy score is given between the ratio of stars with a retrieved period laying between $\pm 10 \%$ error according to the right period over the total number of stars). Our 14,441 stars are distributed between a training set of 12,275 stars (85 \%) and a test set of 2,166 stars (15 \%). The distribution is randomly chosen. To check whether it could be responsible for a bias in the training, we compute ten trainings of the algorithm with different distributions each time. The average classifier accuracy and true accuracy over those ten runs are respectively 0.936 and 0.979 (see Figure~\ref{fig:second_rf} for an example of the training).

\section{Conclusions}

Random forest classifiers prove themselves to be an excellent tool to study stellar rotation properties and allow us to deal with large datasets. On the two distinct steps of the analysis, we get promising accuracy values of 0.895 for the rotators classification and 0.979 for the retrieval of rotation period. Especially, the classifier seems particularly efficient to retrieve the filter that leads to the rotation period. However, we still need to improve the accuracy of the results in the future, especially by using larger data sets. One of the goal of future work will be to apply the analysis to datasets from other missions like K2 or TESS.

\begin{acknowledgements}

 This paper includes data collected by the \textit{Kepler} mission and obtained from the MAST data archive at the Space Telescope Science Institute (STScI). Funding for the \textit{Kepler} mission is provided by the NASA Science Mission Directorate. STScI is operated by the Association of Universities for Research in Astronomy, Inc., under NASA contract NAS 5–26555. This work has been partially fund by the GOLF and PLATO grants at the CEA. A.R.G.S acknowledges the support from National Aeronautics and Space Administration (NASA) under the grant NNX17AF27G. S.M. acknowledges the support from the Ramon y Cajal fellowship number RYC-2015-17697. S.N.B. thanks all the SSEBE team at the IAC for the all the scientific discussions and support during the internship period at the IAC.

\end{acknowledgements}

\bibliographystyle{aa}  
\bibliography{breton_S15} 

\section*{A1 Random forest classifiers}

\begin{figure}[ht!]
    \centering
    \includegraphics[width = 0.4 \textwidth]{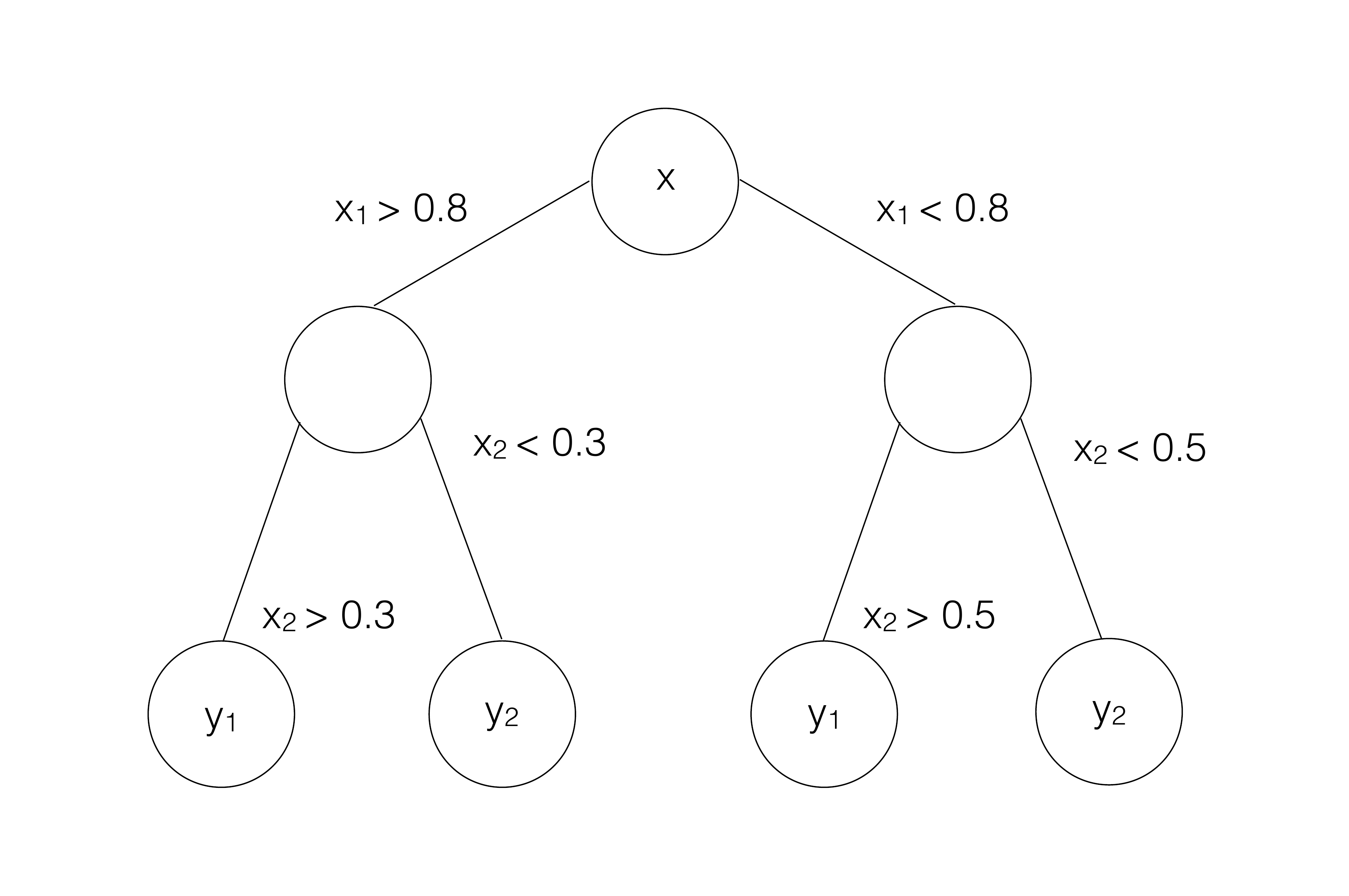}
    \caption{Example of decision tree designed to classify two-parameters data $\bm{x} = \{x1, x2\}$ within two classes $\bm{y}_1$ and $\bm{y}_2$.}
    \label{fig:decision_tree}
\end{figure}

A random forest algorithm is a useful machine learning tool for classification. Thanks to the training set, the algorithm is able to grow a \textit{forest} of decision trees that will then be used to assign a label to new data. A simple example of decision tree is showed in Figure~\ref{fig:decision_tree}

Decision trees are built with the following principle. The training data set is split from the root of the tree according to the value of one of the parameters of the data. Each resulting node gets a Gini score G :

\begin{equation}
        \mathrm{G} = \sum\limits_{k=1}^{\mathrm{N_{\mathrm{classes}}}} p_k \times (1-p_k) ;
\end{equation}
that quantifies its purity (e.g. the proportion of each class for the data assigned to the node). A Gini score of 0 means that a node is totally pure (i.e. that all the elements assigned to the node have the same class). When the score of a node is low enough, the splitting stops: the node becomes a leaf that assignates to new data the label of the dominant class. 

In order to choose a split close to the optimal possibility without consuming too much computation time, a number of possible splits is randomly generated and the best one is chosen over this sample. 

\end{document}